\documentclass[a4paper,11pt]{article}
\usepackage{pos}
\usepackage{natbib}

\title{Pushchino multibeams pulsar search. First results.}
\author*[a]{Sergey A. Tyul’bashev}
\author[b]{Gayane E. Tyul’basheva}
\author[a]{Marina A. Kitaeva}

\affiliation[a]{Lebedev Physical Institute, Astro Space Center, Pushchino Radio Astronomy Observatory,\\
Radiotelescopnaya 1a, Moscow reg., Pushchino, 142290, Russia}

\affiliation[b]{Institute of Mathematical Problems of Biology, brunch of Keldysh Institute of Applied Mathematics,\\
Vitkevich 1, Moscow reg., Pushchino, 142290, Russia}

\emailAdd{serg@prao.ru}
\emailAdd{g.tyulbasheva@yandex.ru}
\emailAdd{marina@prao.ru}

\abstract{
Since the discovery of pulsars, dozens of surveys have already been conducted with their searches. In the course of surveys in the sky, areas from thousands to tens of thousands of square degrees are explored. Despite repeated observations of the same areas, new pulsars are constantly being discovered.  We present Pushchino Multibeam Pulsar Search (PUMPS), having a sensitivity that is an order of magnitude higher than the sensitivity of all previously made surveys on pulsar search. In PUMPS daily round-the-clock observations are carried out of the area located on declinations $-9^o < \delta < +42^o$. The survey is carried out on 96 beams of a Large Phased Array  (LPA) at a frequency of 111 MHz. During the observation period of August 2014 – August 2022, the survey was repeated approximately 3,000 times. The expected sensitivity in the survey reaches up to 0.1 mJy. The paper considers some tasks that can be solved when processing the received data.
}

\FullConference{%
  The Multifaceted Universe: Theory and Observations -  2022 (MUTO2022)\\
  23-27 May 2022\\
  SAO RAS, Nizhny Arkhyz, Russia\\}

\begin{document}
\maketitle

\section{Introduction}

The surveys on the pulsar search are conducted since their discovery in 1967  \cite{Hewish1968} and they have already led to the discovery of more than 3,300 pulsars (https://www.atnf.csiro.au/ research/pulsar/psrcat/, \cite{Manchester2005}). The estimate of the expected number of observed radio pulsars is 30,000 pulsars with a luminosity higher than 0.1 mJy per kpc$^2$ (\citeauthor{Lorimer2006}, \citeyear{Lorimer2006}). The estimate of the number of pulsars available for observations on the radio telescope Square Kilometer Array (SKA) under construction is 20,000 pulsars (\citeauthor{Cordes2004}, \citeyear{Cordes2004}). That is, about 10-15\% of the radio pulsars available for observation have been discovered so far.

In the paper \citeauthor{Wilkinson2007} (\citeyear{Wilkinson2007}), it was shown that the number of new major discoveries in pulsar searches increases as a natural logarithm of the number of discovered pulsars. For each subsequent discovery, it is necessary to discover many times more pulsars than they were known at the time of the previous discovery. A natural question arises about the point of conducting new surveys. Having more and more time and financial expenditures for conducting the observations themselves and processing them, the experimental scientist has less and less chance for a major discovery.

Surveys conducted with high sensitivity make it possible not only to discover new types of pulsars (\citeauthor{McLaughlin2006} (\citeyear{McLaughlin2006}), \citeauthor{Caleb2022} (\citeyear{Caleb2022})), but also to study pulsar populations in detail, to explore the interstellar medium both in the Galactic plane and in the halo. Therefore, if expenditures of time and other resources are acceptable, the surveys should be carried out, because they improve our knowledge about the evolution of pulsars and their properties. 

Surveys are conducted, as a rule, on an area of the sky accessible to the telescope. That is, on about half of the celestial sphere. Taking into account the small, as a rule, dimension of the antenna pattern and the number of simultaneously available beams, the survey is an extremely long and, taking into account the cost of an hour of observations, also an extremely expensive task for almost all large telescopes. However, for the Large Phased Array (LPA) radio telescope located in Pushchino Radio Astronomy Observatory (PRAO), a survey of the entire sky is a daily routine task. In this paper, we consider PUshchino Multibeams Pulsar Search (PUMPS) and some of the problems that can be solved along the way in this survey. 

\section{Survey and tasks}

The survey on the LPA transit radio telescope (observation frequency 111 MHz) is carried out daily, around the clock since August 2014 on 96 beams, and since January 2022  on 128 beams aligned in the meridian plane and covering declinations  $-9^o < \delta < +55^o$. The data are recorded in the 2.5 MHz band, in 32 frequency channels of 78 kHz width  with a point reading time of 12.5 ms. The amount of data recorded per year is almost 45 terabytes. Work on the pulsar search started in 2015, and to date 42 pulsars and 46 rotating radio transients (RRAT) have been discovered up to date (see site https://bsa-analytics.prao.ru/en/ and references in it). Since we have not had  a good computation server, processing of all the data was impossible. We expect that the new server being purchased, which has a terabyte of RAM and 128 full-fledged cores, will start operation this year and will allow processing in a reasonable time the accumulated data with a volume of about 300 terabytes. 

\begin{figure*}
\begin{center}
	\includegraphics[width=0.9\textwidth]{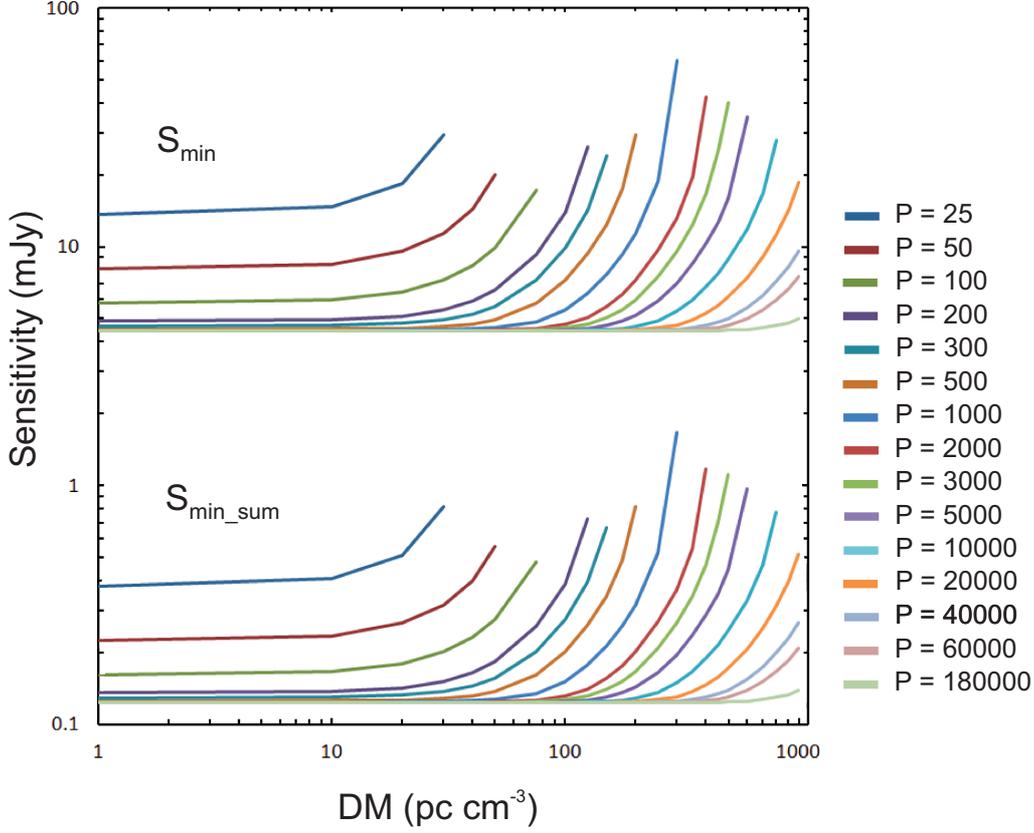}
    \caption{Sensitivity when searching for pulsars (vertical axis) depending on $DM$ (horizontal axis) for different periods of pulsars. In the upper part of the picture, there is the sensitivity $S_{min}$, in the lower part of the picture, there is the sensitivity $S_{min-sum}$. The colors on the right indicate the periods in ms displayed by the curves.}
    \label{fig:fig1}
\end{center}
\end{figure*}

Based on PUMPS data, the search for classical second-duration pulsars and transients is carried out. As was shown in the paper \citeauthor{Tyulbashev2022} (\citeyear{Tyulbashev2022}), the sensitivity of the LPA in one 3.5 minutes duration observation session, when the source passes through the meridian at half the power of the radiation pattern, is inferior to the surveys conducted on the aperture synthesis Low Frequency Array (LOFAR) system and on the Five-hundred-meter Aperture Spherical Telescope (FAST).

Accumulation of the signal by summing up power spectra and periodograms allows to improve the sensitivity by tens of times on the dispersion measures $DM<100$ pc/cm$^3$. For the search for transients, instantaneous sensitivity is primarily important, and this sensitivity is provided by a large effective area of the LPA equal to 45,000 sq.m.

In the paper \citeauthor{Tyulbashev2022} (\citeyear{Tyulbashev2022}), we considered sensitivity when searching for seconds duration pulsars and limited ourselves to PUMPS sensitivity estimates $DM \le 200$ pc/cm$^3$. However, the discovery of the pulsar with a period of 77 seconds and a pulse half-width of 300 ms (\citeauthor{Caleb2022} (\citeyear{Caleb2022})) allows us to seriously consider the search for pulsars at significantly larger $DM$. For such $DM$, the main factor reducing the sensitivity of the search is interstellar scattering ($\tau_s$). In experimental dependencies $\tau_s$(DM) it can be seen that for the same DM the scattering can differ by three orders of magnitude (\citeauthor{Cordes2002} (\citeyear{Cordes2002}), \citeauthor{Bhat2004} (\citeyear{Bhat2004}), \citeauthor{Kuzmin2007} (\citeyear{Kuzmin2007}), \citeauthor{Pynzar2008} (\citeyear{Pynzar2008})). So, in observations on the frequency 111 MHz the scattering can be from ten seconds to ten minutes. We have recalculated the sensitivity curves up to $DM=1,000$ pc/cm$^3$. The following formula was used for estimating the scattering (\citeauthor{Cordes2002}, \citeyear{Cordes2002}):

\begin{equation}
\log(\tau_s) = 3.59 + 0.129\log(DM) + 1.02\log(DM)^2 - 4.4log(f),
    \label{eq:1}
\end{equation}
where $\tau_s$ is obtained in microseconds, $DM$ is expressed in pc/cm$^3$, $f$ is the central frequency of observations in GHz. Since the estimates of $\tau_s$  may be significantly higher than those obtained from formula~\ref{eq:1}, this may lead to a deterioration in the sensitivity assessment both in single observation sessions ($S_{min}$) and when summing up power spectra and periodograms ($S_{min-sum}$).

Fig.~\ref{fig:fig1} shows sensitivity estimates for pulsars with different periods after evaluation of  $\tau_s$ using formula~\ref{eq:1}. The sensitivities in a pulsar search with different periods and dispersion measures shown in the figure differ slightly from the sensitivities shown in fig.4 in the paper \citeauthor{Tyulbashev2022} (\citeyear{Tyulbashev2022}). These differences are related to the fact that in the paper \citeauthor{Tyulbashev2022} (\citeyear{Tyulbashev2022})2 the scattering was estimated according to the empirical formula from the paper \citeauthor{Kuzmin2007} (\citeyear{Kuzmin2007}).

The sensitivity in PUMPS, equal to 0.2-0.3 mJy, is about 16 times better than sensitivity equal to 3-4 mJy in the survey LOTAAS made on LOFAR (Sanidas2019), when recalculated into the frequency 111 MHz (\citeauthor{Tyulbashev2022} (\citeyear{Tyulbashev2022})). This means that, all other things being equal, 4 times weaker pulsars can be detected on LPA than on LOFAR. In addition, there is a possibility in principle to find pulsars with long (>10-20 s) periods at high DM. Since the luminosity decreases in proportion to the square of the distance, and the volume increases in proportion to the cube of the distance, the number of pulsars available in the survey can grow up to 64 times.  There are 73 pulsars discovered in the LOTAAS survey, and the possible number of new pulsars in the PUMPS survey may reach $64 \times 73 \approx 4,500$. This fantastic assessment is most likely very far from reality. The sensitivities indicated in Fig.~\ref{fig:fig1} are achieved only for pulsars that do not have pulse gaps and with very stable (not variable) radiation.  Let us note, however, that when we conduct the search, the sensitivity is several times lower than expected with the accumulation of 8 years of observations (see site https://bsa-analytics.prao.ru/en/,  references in it and the paper \citeauthor{Tyulbashev2022} (\citeyear{Tyulbashev2022})), for the same areas in the sky, we detect \textbf{all} pulsars discovered on LOFAR and detect almost the same number of new pulsars in these areas that were not detected in the LOTAAS survey  (see Fig.6 in the paper \citeauthor{Tyulbashev2022} (\citeyear{Tyulbashev2022})). Our conservative scenario is the discovery of 200-300 new pulsars in PUMPS, the optimistic scenario is the discovery of 1,000-1,500 new pulsars. 

The sensitivity in the search for pulsed radiation of transients for the LPA radio telescope is fixed, because we cannot change neither the area of the antenna, nor the temperature of the system, nor the band. However, for a year of observations, taking into account the 3.5-minute passage through the meridian at half power, approximately 20 hours of recording are accumulated for each point in the sky. The survey started in August 2014 and is planned at least until December 2024. This means the accumulation of approximately 8.5 days of data for each point entering the observation area. In the papers \citeauthor{Logvinenko2020} (\citeyear{Logvinenko2020}), \citeauthor{Tyulbashev2022a} (\citeyear{Tyulbashev2022a}), the existence of RRAT is shown, between the pulses of which 10 or more hours can pass.  To detect and study such transients, it is necessary to conduct very long-term observations, which appear during monitoring.

The field of view of the LPA on 128 antenna beams is approximately 50 sq.deg.. Field of view estimates of other large telescopes used for RRAT detection are:  for 64-meter mirror Parks (Australia) it it is 0.7 sq.deg. on 13 beams at the frequency 1.4 GHz; for 100-meter mirror Green-Bank (USA) it is 0.35 sq.deg. on one beam at the frequency 350 MHz; for 300-meter mirror Arecibo (USA) it is 1 sq.deg. on 7 beams at the frequency 327 MHz; for 500-meter mirror FAST (China) it is 0.16 sq.deg. on 19 beams at the frequency 1.2 GHz. Instant sensitivity for FAST (\citeauthor{Han2021}, \citeyear{Han2021}) after recalculation of the frequency 1.2 GHz into the frequency 111 MHz with an assumed spectral index 1.7 ($S\sim\nu^{-\alpha}$) exceeds the sensitivity of the LPA by about an order of magnitude. However, if we talk about RRAT, for which hours can pass between the appearance of successive pulses, the second main factor for searching, after instantaneous sensitivity, becomes the time of observations at one point in the sky. If the average time between RRAT pulses is one hour, then the FAST radio telescope will need to view half of the sky once [20,000 sq.deg. (half of the celestial sphere)/0.7 sq.deg. (FAST field of view )] $\times$ 1 hour = 28,570 hours or 3.3 years of round-the-clock observations. Due to the  FAST availability, even for a single examination of the sky, the task looks hardly realizable. 

Thus, both for the search for second-duration pulsars and for the search for RRAT, the LPA radio telescope turned out to be surprisingly suitable, despite all its obvious disadvantages: observations in one linear polarization (whole classes of tasks fall out plus a loss of sensitivity 2$^{1/2}$ times); narrow full band (leads to low accuracy of DM estimation and deterioration of sensitivity obtained on modern broadband recorders); the lack of direct ascension diagram control (leads to low accuracy in determining the pulsar period at an interval of 3.5 minutes, there are problems in obtaining timing, it is impossible or very difficult to investigate weak sources); the dimension of one LPA beam is too large $0.5 \times 1$ deg. (it leads to low coordinate accuracy of detected objects).

Despite the mentioned disadvantages of the tool, the data obtained at the LPA can be used for research for many scientific tasks. We list some of the planned PUMPS tasks without going into details of their solutions. Search tasks: search for pulsars with periods from 25 ms up to minutes, search for RRATs and Fast Radio Bursts (FRBs), search for pulsars in nearby galaxies, search for pulsars with small DM down to 0 pc/cm$^3$, search for pulsars with sporadic radiation.

Research of the interplanetary, interstellar, intergalactic environments: pulsar variability induced by scintillations (diffraction and refraction of radio emission in different medium), pulse scattering of pulsars and FRB, Faraday rotation. Research of pulsed and periodic radiation sources: a nature of RRAT and FRB, statistics of pulsars with inter-pulses, inter-pulse radiation of pulsars, pulsars with fading radiation, intrinsic variability, targeted search for gamma, X-ray and other radio-quiet pulsars, spatial distribution of pulsars in the Galaxy as a whole and for different samples, pulse energy distribution at a frequency of 111 MHz, timing and others.

Let's look at three examples of using monitoring data: 

- there are opposite hypotheses about the evolution of the pulsar's "magnetic axis" relative to its axis of rotation. There are hypotheses according to which, over time, the directions of the pulsar's "magnetic axis" and its axis of rotation become perpendicular to each other (orthogonal rotator), other hypotheses suggest that the direction of the axes coincides with time (coaxial rotator) (see \citeauthor{Arzamasskiy2017} (\citeyear{Arzamasskiy2017}) and references there). For the orthogonal rotator, the pulse takes the smallest possible fraction of the period, for coaxial rotators, on the contrary, the pulse takes the maximum possible fraction of the period. Pulsars with long periods are old pulsars (see, for example, the hand book \citeauthor{Lorimer2004} (\citeyear{Lorimer2004})). Their evolution took longer time compared to ordinary second-duration pulsars. Therefore, a simple comparison of the relative pulse duration (duty cycle), that is, the fraction of the period occupied by the pulse for ordinary second-duration pulsars and for pulsars with extra-long periods (>10-20 s), which will be discovered in the survey, should give an answer to the question whether radio pulsars become coaxial or orthogonal rotators by the end of their life in the active phase;

\begin{figure*}
\begin{center}
    \includegraphics[width=0.75\textwidth]{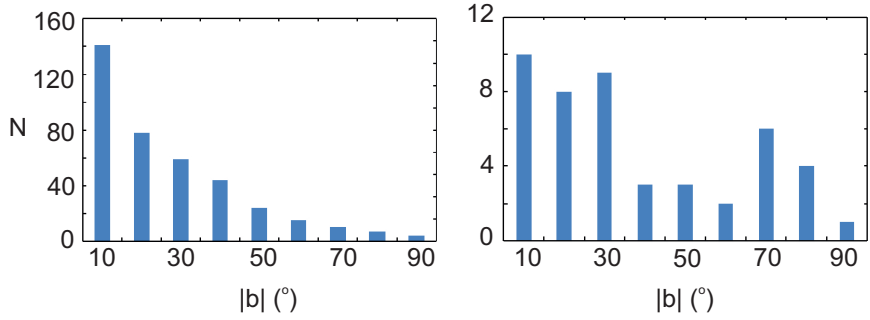}
    \caption{Histograms show the dependence of the observed number of pulsars on the galactic latitude. The left part of the figure shows the dependence for second-duration pulsars from ATNF with $P>0.4$ s, $DM<100$ pc/cm$^3$, $-9^o < \delta < +55^o$. The right part of the figure shows dependence for rotating radio transients detected in PUMPS. The vertical axis shows the number of pulsars in the cells of the histogram, the horizontal axis shows the galactic latitude. The dimension of the histogram cell is equal to $10^o$. For negative latitudes, their modulus was used}
    \label{fig:fig2}
\end{center}
\end{figure*}

- since pulsars are formed during supernova explosions, and the exploded stars are in the plane of the Galaxy, then pulsars at birth should be located there as well. As a result of the explosion, pulsars can acquire a velocity component perpendicular to the plane of the Galaxy and go into the halo. Pulsar lifetime in the active phase (as a radio pulsar) can be from millions to tens of millions of years, and the pulsar can move away from the plane of the Galaxy by some distance. However, the farther away from the Galactic plane, the fewer pulsars should be detected. Fig.~\ref{fig:fig2} presents histograms showing the number of pulsars and RRATs at different galactic latitudes (at different elevations above the plane of the Galaxy). It is obvious that the distributions for second-duration pulsars with small DM located in the same area where all Pushchino RRATs were detected (see https://bsa-analytics.prao.ru/en/transients/rrat/), and the distribution for Pushchino RRATs are  different in appearance. We do not discuss this difference, which can be explained within the framework of hypotheses from insufficient RRAT statistics and selection effects to the discovery of relic pulsars inherited from the previous Universe (\citeauthor{Gorkavyi2021}, \citeyear{Gorkavyi2021}). We are only presenting here one of the problems associated with the strange dependence of the RRAT distribution over Galactic latitudes;

\begin{figure*}
\begin{center}
    \includegraphics[width=0.9\textwidth]{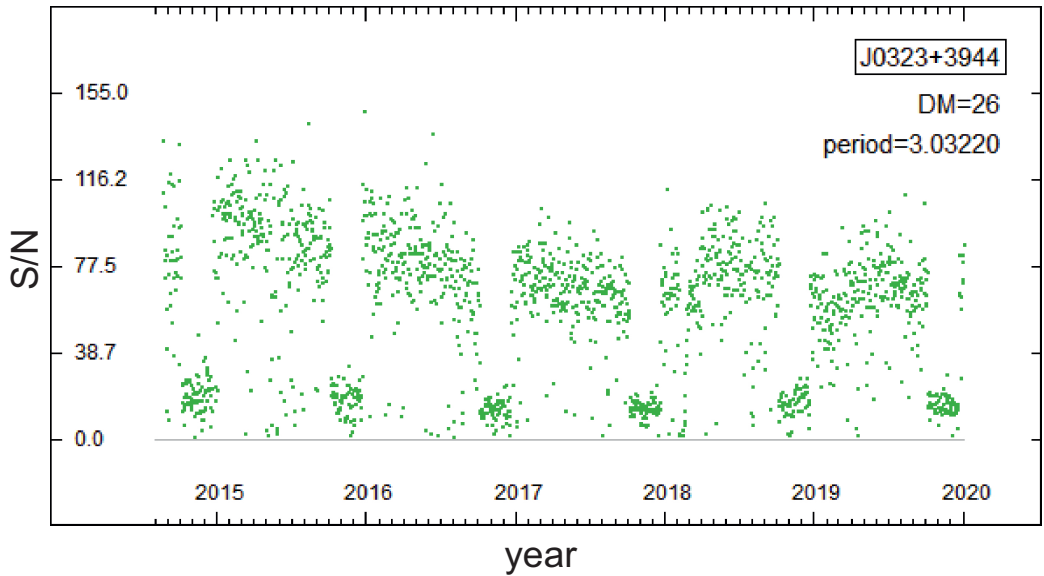}
    \caption{The figure shows the height estimate in the averaged profile for J0323+3944 on time. Each point is an estimate for one day. The vertical axis shows the value of $S/N$, on the horizontal axis shows the date of observations.
The figure shows changes in the flux density over time.}
    \label{fig:fig3}
\end{center}
\end{figure*}

- daily observations allow us to obtain an averaged profile for hundreds of pulsars. The peak or integral flux density estimated from the average profile allows us to construct the dependence of the estimate of the observed flux density on time. The observed variability may be related to scintillations on interstellar plasma. If the variability cannot be explained by the interstellar medium, it is related to internal factors. Fig.~\ref{fig:fig3} shows the "light curve" of the pulsar J0323+3944 (B0320+39). The figure shows changes in the flux density over time. We do not investigate in this paper the causes of apparent variability of J0323+3944, but we only show that the task of studying the variability can be solved on the PUMPS data.

In the presented paper there are no solutions of the problems discussed above in the examples, this is a matter of the future. We only show the fundamental possibility of performing various tasks based on the data received in PUMPS.

\section{Conclusion}

Up to date, 88 pulsars have been discovered in the PUMPS survey (see cite https://bsa-analytics.prao.ru/en/). Having a sensitivity an order of magnitude higher than in the surveys conducted up to date, we can expect the detection of more than 1,000 new pulsars. The main thing, in our opinion, is that with a radical increase in sensitivity, we begin to exploit the area that is called "unknown-unknown" in the paper \citeauthor{Wilkinson2015} (\citeyear{Wilkinson2015}). When working in this area, there is no guarantee that major discoveries will be made, however, as we believe, this kind of work is the real academic science.

\acknowledgments
The study was carried out at the expense of a grant Russian Science Foundation 22-12-00236, https://rscf.ru/project/22-12-00236/.

\end{document}